\documentclass{emulateapj}

\shorttitle{Deep {\it Chandra} Observation of A2052}
\shortauthors{Blanton, Randall, Douglass, Sarazin, Clarke, \& McNamara}

\slugcomment{Astrophysical Journal Letters, in press}

\begin{document}

\title{Shocks and Bubbles in a Deep {\it Chandra} Observation of the Cooling Flow Cluster Abell~2052}

\author{E. L. Blanton\altaffilmark{1},
S. W. Randall\altaffilmark{2},
E. M. Douglass\altaffilmark{1},
C. L. Sarazin\altaffilmark{3},
T. E. Clarke\altaffilmark{4}
and B. R. McNamara\altaffilmark{5,2,6}}

\altaffiltext{1}{Institute for Astrophysical Research,
Boston University, 725 Commonwealth Avenue, Boston, MA  02215;
eblanton@bu.edu, emdoug@bu.edu}

\altaffiltext{2}{Harvard Smithsonian Center for Astrophysics,
60 Garden Street, Cambridge, MA 02138;
srandall@head.cfa.harvard.edu}

\altaffiltext{3}{Department of Astronomy, University of Virginia,
P. O. Box 400325, Charlottesville, VA  22904-4325;
sarazin@virginia.edu}

\altaffiltext{4}{Naval Research Laboratory, 4555 Overlook Avenue SW, 
Washington D. C. 20375 and Interferometrics, Inc.,
13454 Sunrise Valley Drive No. 240, Herndon, VA  20171; tracy.clarke@nrl.navy.mil}

\altaffiltext{5}{Department of Physics and Astronomy, University of Waterloo,
Waterloo, ON N2L 2G1, Canada;
mcnamara@sciborg.uwaterloo.ca}

\altaffiltext{6}{Perimeter Institute for Theoretical Physics, 31 Caroline St., N. Waterloo,
Ontario, Canada, N2L 2Y5}

\begin{abstract}
We present results from a deep {\it Chandra} observation of Abell 2052.  A2052 is a bright, nearby, cooling flow cluster, at a 
redshift of z=0.035.  Concentric surface brightness discontinuities are revealed in the cluster center, and these features are 
consistent 
with shocks driven by the AGN, both with Mach numbers of approximately 1.2.
The southern cavity in A2052 now appears to be split into two cavities with the southernmost cavity 
likely representing a ghost bubble from earlier radio activity.  There also appears to be a ghost bubble present to the NW of the 
cluster center.  
The cycle time measured for the radio source is $t\approx2\times10^7$ yr using either the shock separation or the rise time of the bubbles.
The energy deposited by the radio source, including a combination of direct shock heating and heating by buoyantly rising bubbles inflated by the 
AGN, can offset the cooling in the core of the cluster.
\end{abstract}

\keywords{
galaxies: clusters: general ---
cooling flows ---
intergalactic medium ---
radio continuum: galaxies ---
X-rays: galaxies: clusters --- galaxies: clusters: individual(A2052)
}

\section{Introduction}
The majority of cooling flow clusters contain powerful radio sources
associated with central cD galaxies.
Initial evidence of radio sources displacing, and evacuating cavities in, 
the X-ray-emitting intracluster medium (ICM) was found with ROSAT 
observations of a few sources including Perseus (B\"ohringer et al.\ 1993), 
Abell 4059 (Huang \& Sarazin 1998), and Abell 2052 (Rizza et al.\ 2000).  
Models predicted strong shocks surrounding the lobes which would appear
spectrally hard in X-rays (Heinz, Reynolds,
\& Begelman 1998).
High-resolution images from {\it Chandra} have revealed many more cases
of radio sources affecting the ICM
by displacing it and creating X-ray deficient ``cavities,'' or ``bubbles.''  

The early {\it Chandra} observations revealed that the 
X-ray-bright rims surrounding the radio sources in cooling flow clusters
were cooler, rather than hotter, than the neighboring cluster gas
(e.g., Hydra A, Nulsen et al.\ 2002;
Perseus, Schmidt et al.\ 2002;
Abell 2052, Blanton et al.\ 2003).  
The bright shells showed no evidence of current strong shocks.
More recently, shocks beyond the bubble rims related to radio source outbursts have been found in
a few clusters including M87/Virgo (Forman et al.\ 2005), Hydra A (Nulsen et 
al.\ 2005a), Hercules A (Nulsen et al.\ 2005b), and MS0735.6+7421 (McNamara et 
al.\ 2005).  These shocks were fairly weak with Mach numbers in the range of
1.2 and 1.7 (see McNamara \& Nulsen 2007 for a review).
Ripple features in the X-ray surface brightness
resulting from the propagation of weak shocks
or sound waves as seen in a long observation of Perseus (Fabian et al.\ 2003, 2006)
may also contribute to heating.
Energy input from buoyantly rising bubbles of
relativistic plasma (e.g., Churazov et al.\ 2002), weak shocks 
(e.g., Reynolds, Heinz, \& Begelman 2001), 
and the propagation of sound or
pressure waves are able to offset cooling.

Abell~2052 is a moderately rich, cooling flow cluster at a redshift of
$z=0.0348$.
A powerful radio source, 3C~317, is hosted by the central cD galaxy, UGC 09799.
Abell~2052 was previously observed in the X-ray with $Einstein$
(White, Jones, \& Forman 1997),
{\it ROSAT}
(Peres et al.\ 1998, Rizza et al.\ 2000),
{\it ASCA}
(White 2000), and 
{\it Chandra} (Blanton et al.\ 2001, 2003).

We present a deep {\it Chandra} observation of Abell~2052, combining the earlier
Cycle 1 data with data from Cycle 6.
This longer observation reveals probable shock features exterior to the bubble rims that
contribute to heating in the cluster center.
We assume $H_{\circ}=70$ km s$^{-1}$ Mpc$^{-1}$, $\Omega_{M}=0.3$, and 
$\Omega_{\Lambda}=0.7$ ($1\arcsec = 0.69$ kpc at $z = 0.0348$) throughout.  Errors
are given at the $90\%$ confidence level unless otherwise stated.

\section{Observation and Data Reduction} \label{sec:data}

Abell~2052 was observed with {\it Chandra} using the ACIS-S detector on 2000 September 3 for a total of
36,754 seconds and on 2006 March 24 for 128,630 seconds.
The events from the 2000 data were telemetered in Faint 
mode and the events from the longer exposure were telemetered in VFaint mode.
The data were processed in the standard manner, using CIAO 4.0 and CALDB 3.4.3.
After cleaning, the total exposure remaining for the two data sets was
162,828 seconds.  Background corrections were made using the blank-sky background fields.

\section{X-ray Image} \label{sec:image}

A slightly smoothed (0\farcs49 Gaussian) image of the central region of
A2052 is shown in Figure 1.  The overall emission is fairly circular with a slight ellipticity
in the N-S direction.  
Two separate surface-brightness discontinuities are visible surrounding the cluster center, and these edges
are seen most clearly to the NE at radii of 
approximately 30 and 45 kpc from the AGN (apparent as a bright point source in the center
of the image).  Depressions in the 
surface brightness in the cluster center have been previously identified as bubbles
inflated by the radio lobes associated with the AGN (Blanton et al.\ 2001, 2003).  To the N, there is
a clear inner bubble, and with the combined dataset, an outer bubble is seen to the NW.
The S bubble now appears to be split into an inner and outer bubble, with the outer bubble
to the SE.  These outer bubbles may be ghost bubbles arising from a previous outburst by the AGN.
The radio lobe emission from the AGN fills the outer bubbles and the spectral index steepens in these regions
(Zhao et al.\ 1993).
Bright shells surround the inner bubbles, and a filament extends from the N bright shell to the AGN.  This was
found previously to be associated with H-$\alpha$ emission (Blanton et al.\ 2001), and this feature is now
revealed to extend much closer to the AGN.

\centerline{\null}
\vskip3.0truein
\includegraphics{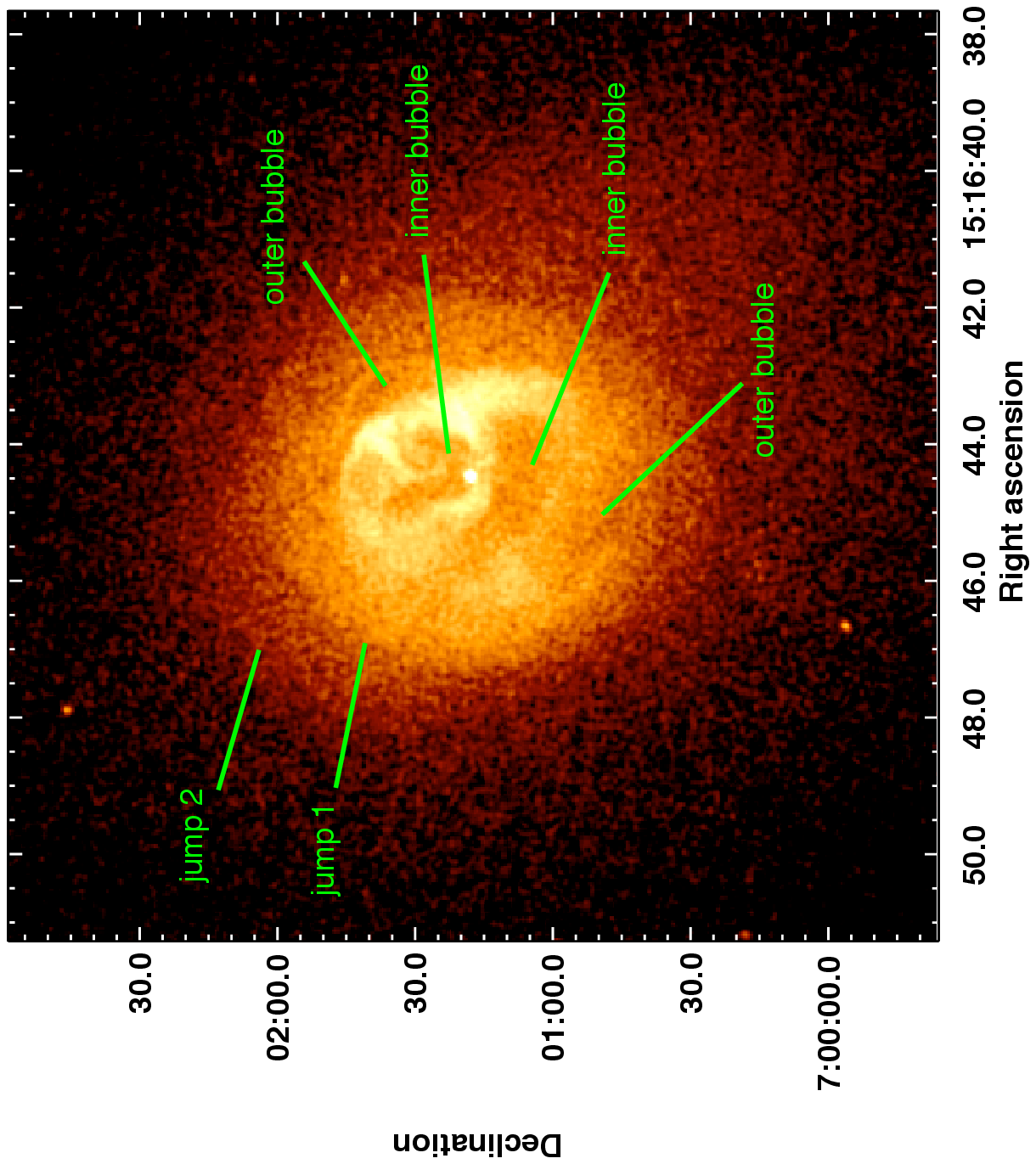}
\figcaption{A 0\farcs49 Gaussian-smoothed {\it Chandra} ACIS-S3 image of the central 3\farcm4$\times$3\farcm4
region of Abell~2052.
The image has been corrected for background and exposure.  Several bubbles related to AGN
outbursts are indicated, as well as discontinuities in the surface brightness surrounding the cluster center.
These discontinuities are modeled as density jumps and are labeled ``jump 1'' and ``jump 2.''}
\vskip0.2truein

An unsharp-masked (0.3--10 keV) image of A2052 is shown in Figure 2 with a slightly larger
field of view than the image shown in Figure 1.  It
was created by dividing an image smoothed with a 0.98 arcsec
Gaussian by an image smoothed with a 9.8 arcsec Gaussian.
The features described above and visible in Figure 1 are also seen in the unsharp-masked image, sometimes
more clearly.  The outer bubbles are revealed more prominently and ripple-like surface-brightness discontinuities are
seen surrounding the cluster center, with two discontinuities visible most clearly to the NE of the AGN.

\section{Density Jump Features} \label{sec:shocks}

To investigate the nature of the surface brightness discontinuities, we have fitted
a spherically symmetric density model, including three power-law components with two 
discontinuities to a NE sector of the cluster and determined the jumps in density that are
represented by these surface brightness changes.  
The NE sector has
an opening angle of 100$^{\circ}$ (position angles $-2^{\circ}$ to $98^{\circ}$ from N).  
The fitted model is shown
with the emission measure profile in Fig.\ 3.  
The emission measure was calculated using projected spectral fits, and is fairly insensitive to temperature.
Assuming a constant temperature, rather than fitting for the temperature in each region, changes the
emission measure values by $\sim5\%$.
The best-fitting model is an excellent fit with
$\chi{^2}=21.2$ for 22 d.o.f.  
A single density jump model does not provide a good fit with $\chi{^2}=116.4$ for 25 d.o.f., nor does a model without
a jump.
With the best-fitting model, two jumps in density are found, at radii of 45 and 67 arcsec (31 and 46 kpc) from the AGN, 
coincident with the surface-brightness edges.  The power-law indices for the three components are
$-0.34,-1.25,$ and $-1.07$, respectively, going from the center of the cluster outward.
The inner density jump is a factor of $1.26^{+0.05}_{-0.03}$ and the outer density jump
is a factor of $1.30^{+0.07}_{-0.05}$.  These density discontinuities are consistent with shocks
with Mach numbers of $1.18^{+0.03}_{-0.02}$ for the 31 kpc feature and $1.20^{+0.05}_{-0.04}$ for the 46 kpc feature.
The inner feature was briefly noted in Blanton et al.\ (2001), and our Mach number is consistent with the limit
found in the earlier paper.
The very similar Mach numbers for the two shocks may indicate that the feedback cycle for A2052 is regular, with similar amounts of 
energy injected into the cluster with each outburst of the radio source.  For each shock, we would expect
the temperature to jump by a factor of 1.2.

\centerline{\null}
\vskip3.1truein
\includegraphics{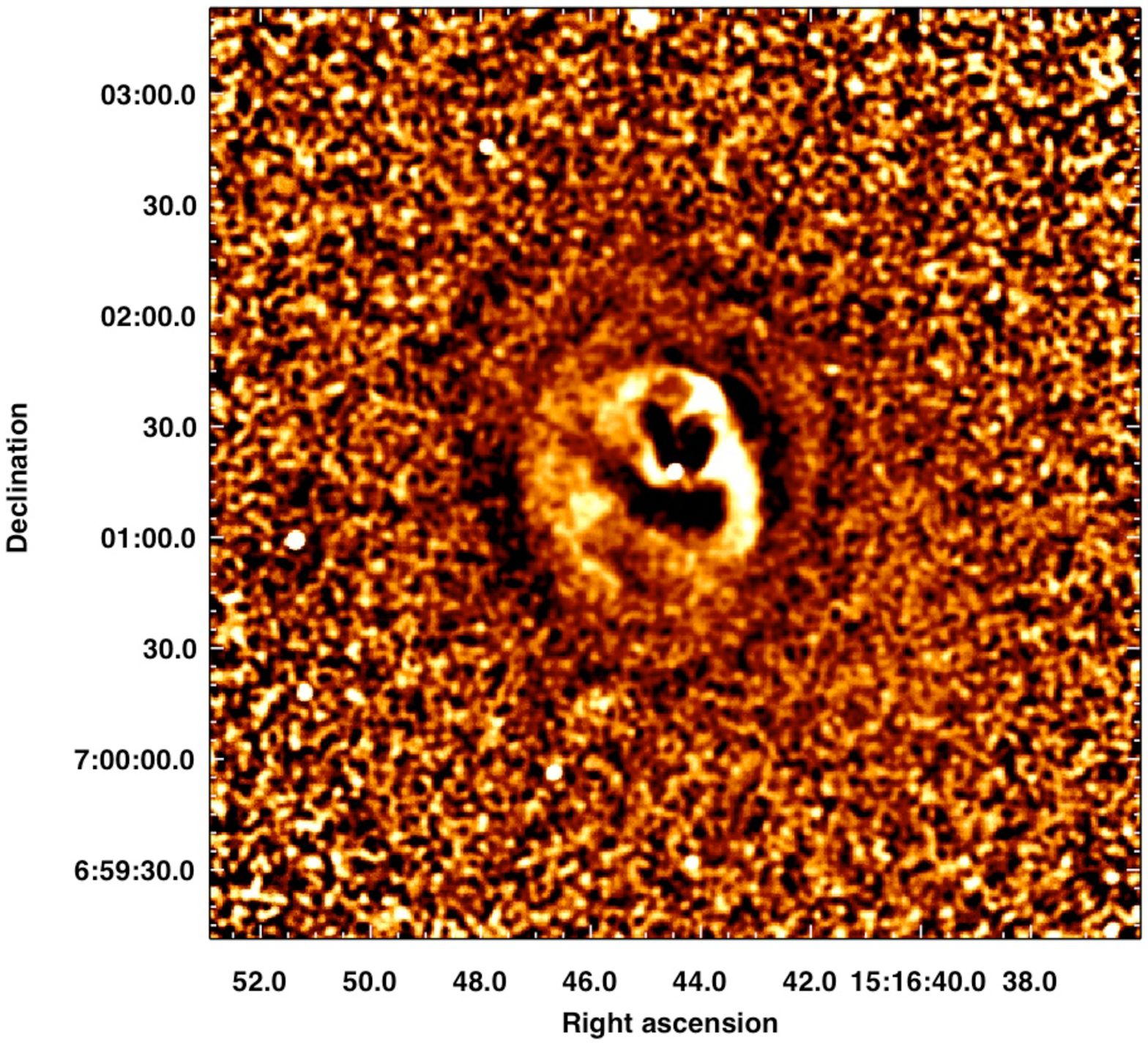}
\figcaption{An unsharp-masked {\it Chandra} ACIS-S3 image of the central 4\farcm2$\times$4\farcm2
region of Abell~2052.
\label{fig:unsharp}}
\vskip0.2truein

\subsection{Spectral Analysis}

In order to examine the temperature structure near the potential shock regions as well as in the cluster center in general,
we generated a temperature map.
The tessellated temperature map is shown in
Figure~\ref{fig:vtmap}.  Each bin was fitted using only counts from that
area, so the extraction regions are well-defined and do not overlap.  The bins were
generated using the algorithm provided by Diehl \& Statler (2006),
which is a generalization of Cappellari \& Copin's (2003) Voronoi
binning algorithm, and requiring roughly 1100 net counts per bin (see Randall et al.\ 2008 for more
details).
Each bin was fitted in the $0.6 - 7.0$ keV energy range
with an APEC thermal plasma model with temperature and abundance allowed
to vary, and absorption fixed at the Galactic value of $N_H = 2.71\times
10^{20}$ cm$^{-2}$.  The color bar gives the best fit temperaures in keV.
No sharp rise in temperature is seen associated with either of the jumps in
surface brightness corresponding to the shock regions.  In general, the cluster is
coolest at the center with the bright shells of emission surrounding the bubbles exhibiting the coolest 
temperatures, $kT \approx 0.8-1.0$ keV.

\centerline{\null}
\vskip2.5truein
\includegraphics{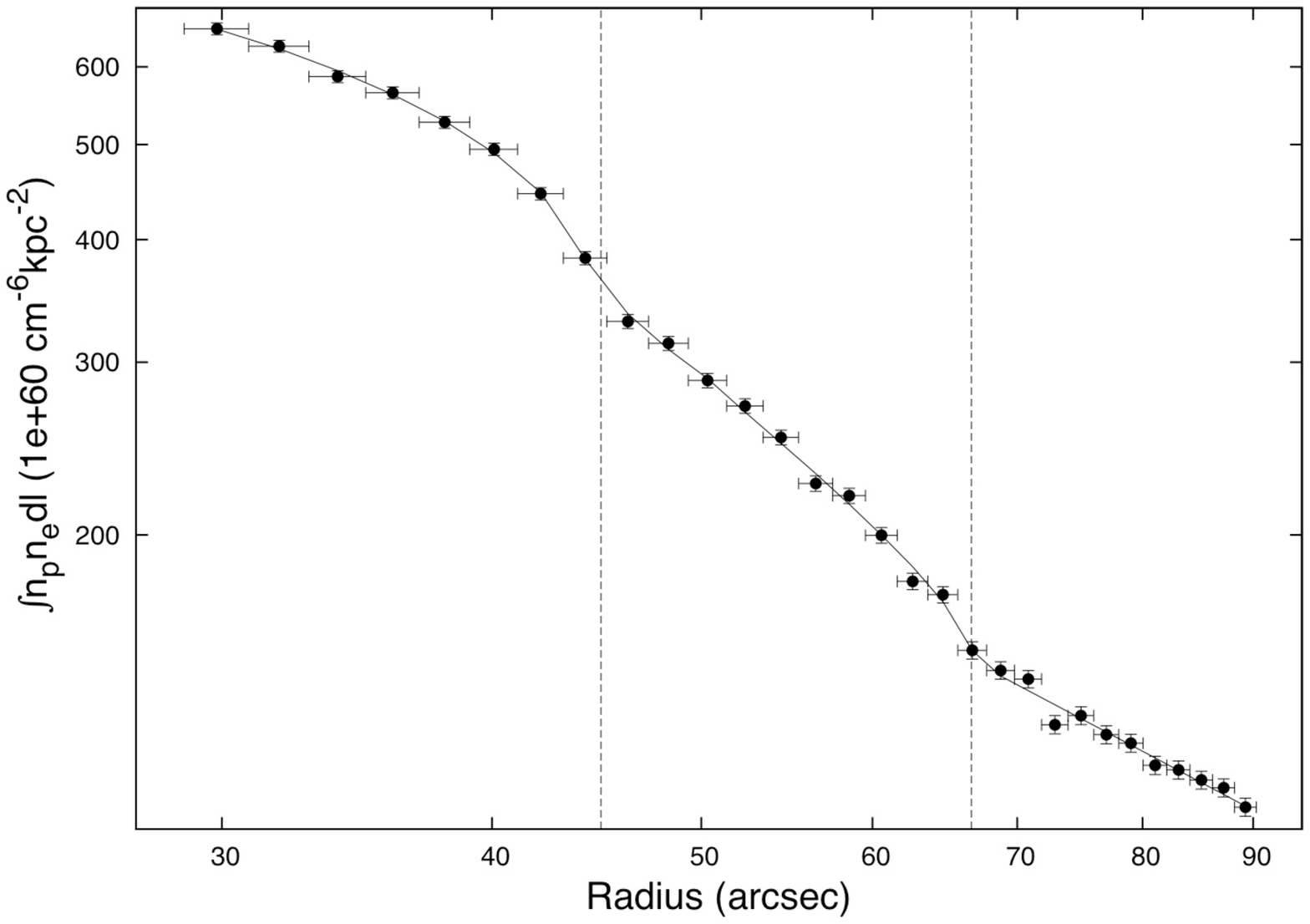}
\figcaption{Emission measure profile for a NE sector of A2052 with a fitted spherically symmetric density model
that includes three power-law components (two discontinuities).  Jumps in density are found at radii of 45 and 67 arcsec (31 and 46 kpc)
from the AGN, as indicated by the dashed lines.
\label{fig:jumps}}
\vskip0.2truein

We have measured the projected temperature profile in the same NE sector used for the emission measure profile.
The fits were performed using the APEC
model, as described above.  The profile is shown as the solid points in the upper panel of Fig.\ 5.  As in the temperature map,
no large jumps in temperature are seen inside with the shocks, indicated with dashed lines in 
Fig.\ 5.
Profiles of density and pressure are also shown
in Fig.\ 5.  The density was determined by deprojecting the surface brightness profile, assuming that the
emissivity is constant in spherical shells.  We use the method described recently in Wong et al.\ (2008),
and also in Kriss et al.\ (1983).
Since the surface brightness is proportional to $n^2T^{1/2}$, the density is not a very sensitive function of 
the temperature.  We have used our projected spectral fits when calculating the conversion from count rate to flux,
which is used in combination with the surface brightness deprojection to determine the density.
As a comparison, we have also calculated the density directly from the deprojected spectral fits, using the projected
volume at each radius, and find that the density values agree to within $\sim5\%$ with those calculated by 
deprojecting the surface brightness.  However, since the errors in the spectral fits (both for the temperatures
and normalizations) are much larger for the deprojected spectra, the errors in the density values are much
larger in this case.

The temperature map in Fig.\ 4 and the temperature profile (solid points) in Fig.\ 5 both display projected temperatures,
and so include the contribution of gas along the line-of-sight to the gas at the radii of interest, in the vicinity of the shocks.
This projection can make it more difficult to detect any temperature rise that may be associated with
the shocks.
We have therefore performed a spectral deprojection for the NE sector.
The spectrum for the outermost annulus was fitted with
a model including Galactic absorption and an APEC model, with temperature and abundance allowed to vary.  The next annulus
in was then fitted with a model including the best fit from the outermost annulus with the normalization scaled to account for
the projection of this outer annulus onto the current annulus being fitted.  Parameters from the outer annulus fit were frozen
and another APEC component was added for the current annulus.  Spectra were fitted in this manner for each annulus, including
the fits from all exterior annuli properly scaled and frozen.  See Blanton et al.\ (2003) for further 
description of this technique.
The temperatures resulting from the deprojection are shown as the open circles in the upper panel of Fig.\ 5.  In addition,
we have used the deprojected temperatures to recalculate the pressures, shown as the open circles in the bottom panel of Fig.\ 5.

With the spectral deprojection, for the inner shock, the best fitting pre-shock temperature is $kT = 2.95^{+0.17}_{-0.18}$ keV
and the post-shock temperature is $kT = 2.97^{+0.30}_{-0.32}$ keV.
For the outer shock, these values are $kT = 3.26^{+0.61}_{-0.44}$ keV and $kT = 3.33^{+0.65}_{-0.44}$ keV, respectively.
Therefore, for the inner shock, the temperatures are consistent with a rise inside the shock by a factor as large as 1.2, and
for the outer shock this value is as large as 1.3.  In both cases, then, the temperatures inside and outside the shocks
are consistent with what is predicted for shocks with Mach number 1.2 (a temperature rise with a factor of 1.2).

\centerline{\null}
\vskip3.3truein
\includegraphics{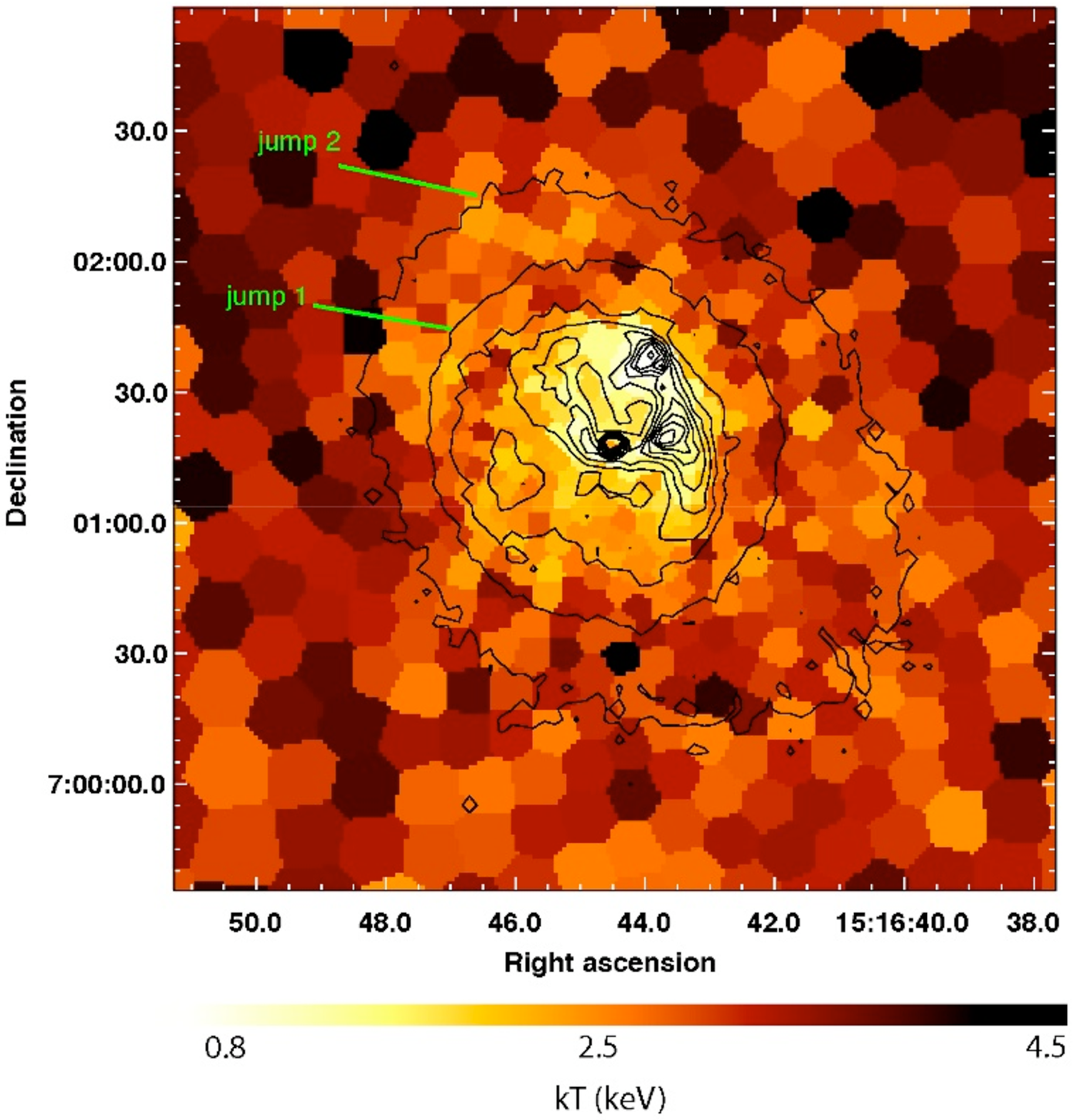}
\figcaption{A Voronoi-Tesselation temperature map of the central region of A2052.  Contours of surface brightness
are superposed.  A large increase in temperature is not seen behind the shock regions in the NE in this projected
map.
\label{fig:vtmap}}

\section{Discussion}

While the temperatures we measure in the density jump regions are consistent, within the errors,  with what is
expected with a shock model, the best-fitting temperatures are approximately constant across the jumps.
If a temperature rise is not seen associated with a potential shock, the abrupt change in density
may be due to an isothermal shock or a cold front.
Possible isothermal shocks have been seen in Perseus (Fabian
et al.\ 2006, Mach 1.26) as well as the clusters A2199 
and 2A 0335+096 (Sanders \& Fabian 2006, both Mach 1.5).  This may
indicate that thermal conduction is efficient in these regions.  However, a lack of an observed rise in temperature may
also result from adiabatic expansion and orientation effects (McNamara \& Nulsen 2007).
Cold fronts typically display a sharp change in density in one direction (see Markevitch \& Vikhlinin 2007 for a review), and are not usually seen
to extend 360$^{\circ}$ around the center of a cluster.  The outer density jump feature is much sharper and less
extended to the NE than to the SW and could be associated with a cold front related to either a merger or 
sloshing of the cD galaxy.  The inner jump feature extends 360$^{\circ}$ around the center of the cluster and is
probably more likely associated with a shock.  However, this geometry could be possible if there were a cold
front occurring along our line-of-sight.  The central cD galaxy has a fairly large peculiar velocity of
$-296\pm90$ km s$^{-1}$ relative to the cluster mean (Oegerle \& Hill 2001), which could be related to merger activity along 
our line-of-sight possibly producing a cold front.

\centerline{\null}
\vskip3.0truein
\includegraphics{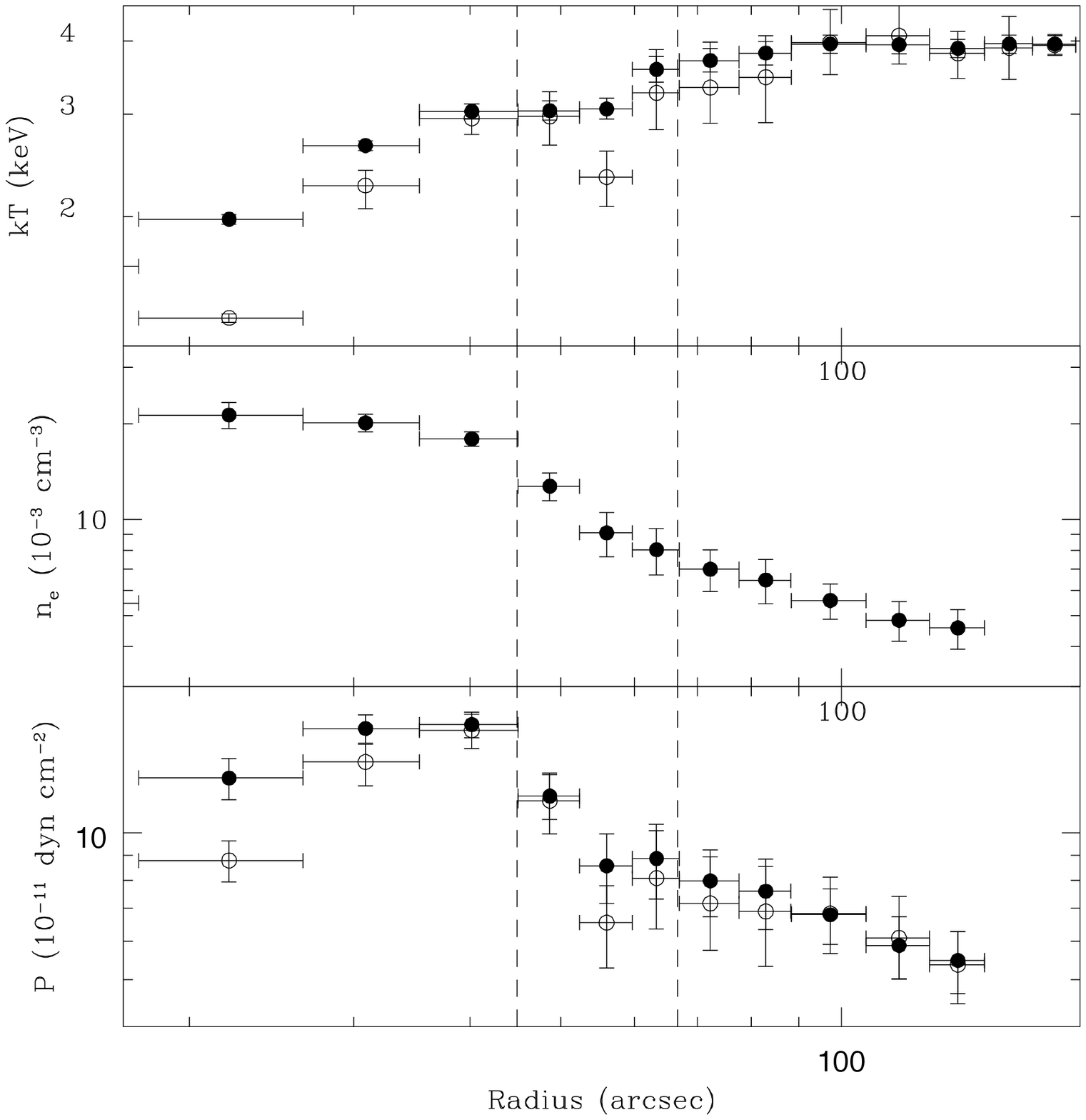}
\figcaption{{\it{Top panel:}}  The projected (filled points) and deprojected (open circles) temperature profiles for the NE sector of the central region of A2052.
{\it{Center panel:}}  Density profile for the NE sector.  {\it{Bottom panel:}}  Pressure profile for the NE sector using the projected (filled points) and deprojected (open circles) temperatures.  Dashed lines in all panels indicate the inner and outer jumps found in the fit in Fig.\ 3.
\label{fig:tprof}}
\vskip0.2truein

We have calculated an average sound speed of 820 km s$^{-1}$ in the central region of the cluster
based on a temperature of
$kT = 2.5$ keV.
If we assume the two density jump features are shocks, we can calculate a cycle time for the radio source (where
the cycle time is the interval between outbursts).
The outer shock is at a projected distance of 46 kpc from the AGN and the estimated speed, based on the
Mach number above, is 980 km/s.
The inner shock is at a projected distance of 31 kpc from the AGN, and the estimated speed for this feature
is 970 km/s.
If the shocks are due to a series of outbursts from the radio
source, their positions and speeds of propagation imply a cycle time of 
$\gtrsim 1.5 \times 10^7$ yr.
A similar cycle time was inferred from the Perseus
data from measuring the ripple separation (Fabian et al.\ 2003).
This time is only slightly longer than the synchrotron radio age of the
lobes in Abell 2052 ($9 \times 10^6$ yr, Zhao et al.\ 1993).  
We can also estimate the cycle time using the locations of
the ghost cavities and assuming that they rose buoyantly at some fraction of the local sound speed.
Using the SE ghost cavity, we estimate a cycle time of approximately $4 \times 10^7$ yr ($2 \times 10^7$ yr)
if the cavities rose at 0.5 times (1 times) the local sound speed,
roughly consistent with our estimate from the shock features.

Fitting a cooling flow model (including Galactic absorption and an extra thermal APEC model to account for overlying gas) 
to the cluster spectra within a radius of 137$\arcsec$ = 95 kpc yields a mass deposition
rate of $\dot{M}=55\pm4~ M_{\odot}$ yr$^{-1}$.  This corresponds to a cooling luminosity of $5.4\times10^{43}$ erg s$^{-1}$.
Heating from the central AGN can come in the form of shock heating as well as
buoyantly rising bubbles inflated by the radio lobes.
If the density jump features represent shocks, we may calculate the shock heating per unit
volume using
\begin{equation}
\Pi_{s} = \frac{(\gamma +1)P}{12 \gamma^2}\left(\frac{\omega}{2\pi}\right)\left(\frac{\delta P}{P}\right)^{3},
\end{equation} 
where P is the pre-shock pressure, $\gamma=5/3$, and $2\pi/\omega$ is the time interval between shocks (McNamara \& Nulsen 2007).
For the inner shock, using the pre-shock pressure of $1.0 \times 10^{-10}$ dyn cm$^{-2}$, and
the expected pressure jump of a factor of 1.55,
we find $\Pi_{s}=2.8 \times 10^{-27}$ erg cm$^{-3}$ s$^{-1}$, or an energy input rate of
$1.0\times10^{43}$ erg s$^{-1}$ within the spherical volume interior to the inner shock using the cycle time
calculated from the potential shock feature separation of $1.5\times10^7$ yr. 
Therefore, energy input from the shocks may contribute to heating, but falls short by a factor of approximately five
of offsetting the cooling.
We may also estimate the energy input from buoyantly rising bubbles.  Here, the heat input comes from
the work done in inflating the bubbles and the energy content of any gas within the bubbles,
$E = (5/2)PV$ (Churazov et al.\ 2002), using $\gamma=5/3$.  Using the pressure in the bubble rims of
$P = 1.1 \times 10^{-10}$ dyn cm$^{-2}$, computing the N and S bubble volumes using a bubble radius of
$10\farcs5$ = 7.2 kpc, and using the cycle time inferred from the outer bubble positions of
$4 \times 10^7$ yr ($2\times10^7$ yr), we find an energy input rate of $2.0\times10^{43}$ erg s$^{-1}$ 
($4.0\times10^{43}$ erg s$^{-1}$).
If we assume that the bubbles are filled with relativistic plasma ($\gamma=4/3$) then $E = 4PV$ (Churazov et al.\ 2002)
and the energy input rate is slightly higher, $3.2\times10^{43}$ erg s$^{-1}$ 
($6.4\times10^{43}$ erg s$^{-1}$).
The combination of direct shock heating and energy input from buoyantly rising bubbles can then offset the cooling
of the gas in the center of A2052.

\acknowledgements
We thank the anonymous referee for comments which significantly improved this paper.
We acknowledge helpful discussions with Christine Jones and Paul Nulsen.
Support for this work was provided by the National Aeronautics and Space
Administration, through {\it Chandra} Award Number
GO5-6137X.  ELB was partially supported by a Clare Boothe Luce Professorship.
Basic research in radio astronomy at the Naval Research Laboratory is supported by 6.1 Base funding.
SWR was supported by the Chandra X-ray Center through NASA contract NAS8-03060.

\end{document}